\documentclass[10pt,preprint,a4paper]{aastex}

\usepackage{amsmath}                
\usepackage{amsfonts}               
\usepackage{amssymb}                
\usepackage{epsfig}                 
\usepackage[left=2.1cm,top=2cm,right=1.8cm,nofoot]{geometry}

\newcommand{\cm}{{~\rm cm}}
\newcommand{\s}{{~\rm s}}

\newcommand{\g}{{~\rm g}}

\newcommand{\erg}{{~\rm erg}}
\newcommand{\yr}{{~\rm yr}}

\newcommand{\AU}{{~\rm AU}}

\newcommand{\days}{{~\rm days}}

\title{A CIRCUMBINARY DISK IN THE FINAL STAGES OF COMMON ENVELOPE AND THE CORE-DEGENERATE SCENARIO FOR TYPE Ia SUPERNOVAE}

\author{Amit Kashi\altaffilmark{1} and Noam Soker\altaffilmark{1}}

\altaffiltext{1}{Department of Physics, Technion -- Israel Institute of
Technology, Haifa 32000 Israel; kashia@physics.technion.ac.il;
soker@physics.technion.ac.il.}

\begin{document}

\begin{abstract}
We study the final stages of the common envelope (CE) evolution and find that a substantial fraction of the
ejected mass does not reach the escape velocity.
To reach this conclusion we use a self-similar solution under simplifying assumptions.
Most of the gravitational energy of a companion white dwarf (WD) is released in the envelope of a massive
asymptotic giant branch (AGB) or the red giant branch (RGB) star in a very short time.
This rapid energy release forms a blast wave in the envelope.
We follow the blast wave propagation from the center of the AGB
outwards, and show that $\sim 1$--$10$ per cent of the ejected envelope remains bound to the remnant binary system.
We suggest that due to angular momentum conservation and further interaction with the binary system,
the fallback material forms a circumbinary disk around the post-AGB Core and the companion WD.
The interaction of the circumbinary disk with the binary system will reduce the orbital separation
much more than expected from the dynamical phase (where the envelope is ejected) of the CE alone.
The smaller orbital separation favors a merger at the end of the CE phase or a short time after, while the core is still hot.
This is another channel to form a massive WD with super-Chandrasekhar mass that might explode as a type Ia supernova.
We term this the core-degenerate (CD) scenario.
\end{abstract}



\section{INTRODUCTION}
\label{sec:introduction}

The 35 years old (\citealt{Paczynski1976}) common envelope (CE) model is in the heart of the formation of many close binary systems
(e.g., \citealt{MeyerMeyer-Hofmeister1979}; \citealt{IbenLivio1993}; \citealt{TaamSandquist2000};
\citealt{Podsiadlowski2001}; \citealt{Webbink2008}; \citealt{TaamRicker2010}).
The CE is defined as a structure where two stars share the envelope.
In most cases the mass of the envelope comes from a giant star as it engulfs its companion, which is either a main-sequence (MS) star or a white dwarf (WD).
During the CE phase the orbital separation decreases due to
gravitational drag and tidal interaction (e.g., \citealt{IbenLivio1993}).
The transfer of orbital energy and angular momentum to the envelope, as well as other possible energy sources, lead
to the ejection of the envelope.

One of the major unsolved questions of the CE phase is the final orbital separation.
It is customary to equate the gravitational energy released by the spiraling-in binary system $E_G$, to the binding
energy of the envelope $E_{\rm bind}$.
An efficiency parameter $\alpha_{\rm CE}$ is introduced such that the final orbital separation is derived from the equality
$E_{\rm bind} = \alpha_{\rm CE} E_G$.
However, different studies deduced different values of $\alpha_{\rm CE}$ for different systems.
\cite{IvanovaChaichenets2011} suggest that the enthalpy rather than the internal energy should be included in calculating the binding energy.
A better understanding of the physical processes that determined the final orbital separation is required.

Most of the energy is deposited into the envelope in a very short time, as the spiraling-in process is
accelerated with the decrease of the orbital separation (e.g., \citealt{RasioLivio1996}; \citealt{LivioSoker1988}).
There are also allegedly contradicting results, which suggest that the time for energy release in the envelope
is relatively long.
For example, \cite{SandquistTaam1998} found that the energy is released
mainly during a stage lasting $\sim 200 \days$ (see their fig. 11).
\cite{DeMarco2003} found that in representative cases the entire timescale for CE evolution is 9--18 years.
However this timescale was obtained by waiting for a negligible amount of material to remain in the envelope \citep{DeMarco2009},
and therefore it represents a much longer timescale than the strong energy release timescale.
This may also suggest that the strong energy release timescale is of the order of $\sim 1$ year or less.
Recent numerical simulations of \cite{Passy2011} also show that the strong energy release occurs in a timescale of $100$--$ 200 \days$.

The energy that is released during the CE is usually composed of more than one burst,
and it takes for the companion a few orbits to strip the asymptotic giant branch (AGB) star envelope away (e.g., \citealt{SandquistTaam1998}).
If most of the energy is released within a short time deep in the envelope, i.e., shorter than the dynamical time
of the outer parts of the envelope, then one can speak of a burst of energy.
Such a burst is similar in some of its properties to a blast wave which propagates outwards to the stellar surface.
This is the scenario we discuss in the paper.

\cite{SandquistTaam1998} showed that some of the mass of the envelope can remain bound to one or both of the interacting stars.
They simulated a $5~\rm{M_{\odot}}$ AGB interacting with a $0.6~\rm{M_{\odot}}$ companion, and found that the
companion unbinds $\sim 1.55~\rm{M_{\odot}}$ ($\simeq 23$ per cent) of the AGB envelope.
\cite{Passy2011} present more extreme results in their simulations.
They find that when the envelope is lifted away from the binary, $\gtrsim 80$ per cent of the envelope remains bound to the binary.
However, in their simulations rotation of the AGB was not included, so the material that remains bound is probably overestimated.
They conclude that in some cases parts of the AGB envelope remains bound or marginally bound to the remnant post-AGB core.
\cite{SandquistTaam1998} also found that a differentially rotating structure resembling a thick disk surrounds the remnant binary
during an intermediate phase of the CE interaction.
This stage occurs briefly before energy deposited by the companion and remnant AGB core drive the mass away.
\cite{Soker1992} (see also \citealt{Soker2004}) have analytically obtained a similar thick disk structure.
\cite{DeMarco2011} suggest that envelope material which is still bound to the binary system at the end of the CE,
will fall back onto the system and will form a circumbinary disk.
They suggest that such a disk might have some dynamical effects on the binary period.
Other simulations show a post-CE gas to be bound to the companion, e.g., \cite{Lombardi2006}.

In this paper we discuss the final stages of the CE under the assumption of a rapid binary-gravitational energy release, and the formation of a circumbinary disk.
We suggest that such a scenario can result in an early merger of the WD with the post-AGB core, that will explode as a Type Ia supernovae (SNe Ia) -- the core-degenerate (CD) scenario.

\cite{SparksStecher1974} developed a merger scenario of a WD with an AGB core.
in which the result is a type II SN.
In a later study, \citet{LivioRiess2003} found that the merger of a WD with an AGB core leads to a SN Ia that occurs at the end of the CE phase or shortly after.
An important conclusion of \citet{LivioRiess2003} is that for  merger to occur the AGB star should be massive.
In the present paper we also reach the same conclusion.

We discuss the problem of the energy budget and characteristic timescale at the end of the CE in section \ref{sec:PROBLEM}.
In section \ref{sec:blast} we discuss the properties of the blast wave.
For such a disk to form some of the material needs to fall-back onto the star.
We follow the blast wave and show that it indeed leaves some of the envelope material bound to the binary system.
We discuss the formation of a circumbinary disk and its potential consequences on the binary system and its final fate.
We argue that in many cases such a disk induces a merger and discuss the implications of the formation of such an early merger (section \ref{sec:implications}).
The most interesting implication is the CD scenario we propose for a SNe Ia.
In the CD scenario a merger of a WD companion with the post-AGB core occurs while the core is still hot, and later explodes as a SN Ia.
In other cases the merger might collapse to a neutron star.
We summarize in section \ref{sec:summary}.

\section{PROBLEM SETUP}
\label{sec:PROBLEM}

\subsection{Energy considerations}
\label{sec:PROBLEM:energy}

In this section we discuss the energy considerations behind the process of expelling the envelope.
We follow a stellar companion spiraling-in inside the AGB envelope, and concentrate on the
orbital separation $a_f$ where the liberated orbital gravitational energy is about equal to the envelope binding energy.
As stated above, the final stages of the spiraling-in occur fast.
Even if the envelope inward to the companion reaches synchronization with the orbital motion,
the spiraling-in will continue.
The reason is that the envelope outer to the companion cannot be synchronized, and tidal interaction are very strong.
This tidal interaction is similar, but not identical, to that with a circumbinary disk.
As we will see in section \ref{sec:implications}, such interactions are very efficient in reducing the orbital separation.
Based on this discussion, we assume that an energy about equal to the binding energy of the envelope is released
in a very short time.
We examine two analytical prescriptions for this energy deposition in section \ref{sec:blast}.

We start by examining the involved energies.
For our typical giant structure we approximate the AGB envelope structures from the models of \cite{NB2006}, \cite{TaurisDewi2004},
and \cite{Soker1992}, that are qualitatively similar.
As we are aiming at performing self similar calculations, we approximate the envelope density profile by a power law, e.g.,
\cite{Soker1992}
\begin{equation}
\rho=\frac{A}{r^\omega},
\label{eq:rho}
\end{equation}
with $\omega=2$.
The mass inward to radius $r$ is given by
\begin{equation}
M(r) \simeq M_{\rm{core}} + M_{\rm{env}}(r) = M_{\rm{core}}
+ \int_{R_{\rm{core}}\simeq 0}^{r} 4\pi r^2 \rho \,dr.
\label{eq:M_r}
\end{equation}
Our AGB stellar mass is $M_\star=5~\rm{M_{\odot}}$, its core mass is $M_{\rm{core}}=0.77~\rm{M_{\odot}}$, 
its envelope mass is $M_{\rm{env}}=4.23~\rm{M_{\odot}}$ and the stellar radius is $R_\star=310~\rm{R_{\odot}}$. 
For these parameters and $\omega=2$ we find from equations (\ref{eq:rho}) and (\ref{eq:M_r})
that $A \simeq 3.1\times10^{19} \g \cm^{-1}$. 
The core radius is very small, few$\times 0.01~\rm{R_{\odot}}$.

\begin{figure}[t]
\resizebox{0.5\textwidth}{!}{\includegraphics{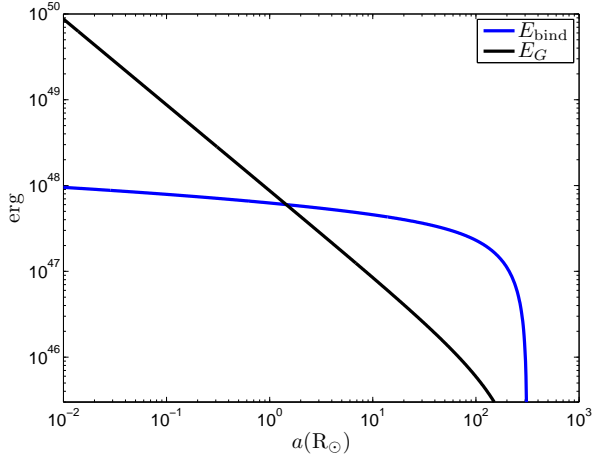}}
\caption{\footnotesize{The binding energy of the AGB envelope mass residing at $r>a$, and the released gravitational energy,
for our case of a $5~\rm{M_{\odot}}$ AGB with envelope density $\sim r^{-2}$ and a $0.6~\rm{M_{\odot}}$ companion, as function of the orbital separation.
As the companion spirals-in towards the center it releases more and more energy,
namely is releases most of its gravitational energy close to the center.
Until it reaches the inner intersection of the binding energy of the AGB and the released gravitational energy
($a_{\rm f} \simeq 1.4~\rm{R_{\odot}}$), the companion liberates $E_{\rm f} \simeq 6 \times 10^{47} \erg$
and unbinds most of the AGB envelope.}}
\label{fig:Ebind_EG_r}
\end{figure}

The binding energy of the envelope mass outside radius $r$ is given by
\begin{equation}
E_{\rm bind} = \int_r^{R_\star} \frac{G(M(r)+M_2)}{r} 4 \pi r^2 \rho dr,
\label{eq:Eb}
\end{equation}
where we have taken into account the companion mass $M_2$, and neglected thermal energy of the gas.
Before the companion enters the envelope its influence is mainly through tidal interaction that
makes the envelope spin-up.
This spin-up and the entrance of the companion to the envelope is likely to cause some mass to be lost.
This mass will be significant if the companion manages to bring the envelope to synchronization with the orbital
motion (\citealt{BearSoker2010}).
We consider a case where if a synchronization is achieved it does so before the AGB reaches its maximum radius.
As the AGB continues to expand it swallows the companion.
We assume that not much mass was lost before the onset of the CE.
The gravitational energy released by the companion inside the envelope as it spirals in from
the stellar surface to binary separation $a$ is given by
\begin{equation}
E_G=\frac{GM(a)M_2}{2a} - \frac{GM_\star M_2} {2 R_\star}.
\label{eq:Eg}
\end{equation}
The binding energy and the released gravitational energy for our AGB model and for $M_2=0.6~\rm{M_\odot}$
are plotted in Fig. \ref{fig:Ebind_EG_r}.
For $a_{\rm f1} \simeq 1.4~\rm{R_\odot}$ the two energies become equal to each other and to $E_{\rm f} \simeq 6 \times 10^{47} \erg$. 
If we use the alpha prescription, $\alpha_{\rm CE} E_G (a_{\rm f \alpha})=E_{\rm bind}(a_{\rm f \alpha})$,
then the ejection of the envelope occurs at an envelope separation of $a_{\rm f \alpha} \simeq \alpha_{\rm CE} a_{\rm f1}$.

\subsection{Timescales}
\label{sec:PROBLEM:times}

In this section we examine some timescales relevant to the final stages of the common-envelope phase.
The spiraling-in timescale due to tidal interaction when the companion is inside the envelope is
about equal to the circularization time
\begin{equation}
\begin{split}
\tau_{\rm{tid}}(a) & \simeq \tau_{\rm{circ}}(a) \simeq 6.9
\left(\frac{L_\star}{2 \times 10^4~\rm{L_{\odot}}}\right)^{-\frac{1}{3}}\\
&\left(\frac{R_\star}{300~\rm{R_{\odot}}}\right)^{\frac{2}{3}}
\left(\frac{M_{\rm{env}(a)}}{4~\rm{M_{\odot}}}\right)^{-\frac{2}{3}} 
\left(\frac{M(a)}{5~\rm{M_{\odot}}}\right)^{-1}\\
&\left(\frac{M_2}{0.2M(a)}\right)^{-1}
\left(1+\frac{M_2}{M(a)}\right)^{-1}
\yr, 
\end{split}
\label{eq:tau_circ}
\end{equation}
where $L_\star$ is the stellar luminosity, and the circularization time is from \cite{VerbuntPhinney1995},
and where inside the envelope we take the radius inward to the companion equals the orbital separation $r=a$.
In this expression only the tidal interaction with the envelope inward to the orbit is considered,
however there is also a tidal interaction with the envelope exterior to the orbit (see section \ref{sec:PROBLEM:energy}), that shortens the process.

We assume that the convection in the envelope inner to the companion
radius still exists.
For that, the tidal dissipation is assumed to be valid.
Spiral wave dissipation exists as well (e.g., \citealt{SandquistTaam1998}),
for which the characteristic timescale is $\sim 200 \days$.

There are two additional timescales to consider.
These are the Keplerian timescale
\begin{equation}
\tau_K(a)=2 \pi \left(\frac{a^3}{GM(a)}\right)^{\frac{1}{2}},
\label{eq:tau_K}
\end{equation}
and the friction timescale (e.g., \citealt{IbenLivio1993})
\begin{equation}
\tau_{\rm{fric}}(a)=\frac{GM(a)M_2}{2aL_{\rm{drag}}(a)}
\label{eq:tau_fric}
\end{equation}
where
\begin{equation}
L_{\rm{drag}}(a) = \xi \pi R_{\rm{acc}}^2 \rho v_{\rm{K}}^3
\label{eq:L_drag}
\end{equation}
is the luminosity due to energy loss by drag,
$v_{\rm{K}}(a) = (GM(a)/a)^{1/2}$ is the Keplerian velocity,
\begin{equation}
R_{\rm{acc}}(a) = \frac{2 G M_2}{v_{\rm{K}}^2}
\label{eq:R_acc}
\end{equation}
is the Bondi-Hoyle accretion radius, and we take the constant $\xi=1$.
The usage of $v_{\rm{K}}$ in the equation (\ref{eq:R_acc}) is an approximation.
A more accurate expression takes into account the rotation of the envelope and the sound speed.
Including envelope rotation reduces the relative velocity, while the sound speed increases it.
Overall, the approximation used here is adequate.

The accretion rate might be less than the Bondi-Hoyle-Lyttleton value.
However, in the drag equation there is another term resulting from interaction of the companion with mass that is not accreted.
This term has the form of (\ref{eq:L_drag}), but with the coefficient $\ln(R_{\rm max}/R_{\rm{acc}})$ instead of $\xi$, where $R_{\rm max}$
is the distance up to where the companion gravity influences the envelope gas (e.g., \citealt{RudermanSpiegel1971}).
This implies that even if the accretion rate is much below the Bondi-Hoyle-Lyttleton value, the term we ignored
will become important and the value of the drag will be about the same as in equation (\ref{eq:L_drag}) with $\xi \simeq 1$.
Therefore, even if there is a large deviation from Bondi-Hoyle-Lyttleton accretion rate,
the value of $\tau_{\rm{fric}}$ will only change by a small factor from the value used here.

\begin{figure}[t]
\resizebox{0.5\textwidth}{!}{\includegraphics{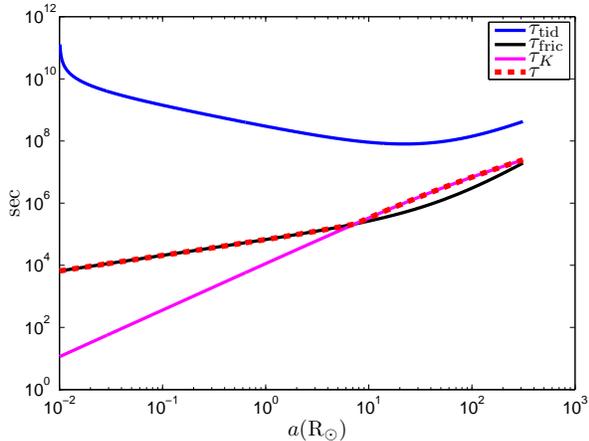}}
\caption{\footnotesize{The characteristic timescales for unbinding the AGB envelope,
for our case of a $5~\rm{M_{\odot}}$ AGB with envelope density $\sim r^{-2}$ and a $0.6~\rm{M_{\odot}}$ companion.
$\tau_{\rm{tid}}$: the spiralling-in timescale due to tidal interaction with the envelope residing inward to the companion orbit.
$\tau_{\rm{fric}}$: the friction timescale.
$\tau_K$: The Keplerian timescale.
$\tau$: the relevant timescale at radius $r$ according to equation (\ref{eq:tau}).
}}
\label{fig:tau}
\end{figure}

The spiraling-in timescale at orbital separation $a$ is dictated by the most efficient process, but under our assumptions it
cannot be shorter than the Keplerian timescale.
It is given by
\begin{equation}
\tau(a)=\max[\min(\tau_{\rm{tid}},\tau_{\rm{fric}}),\tau_K].
\label{eq:tau}
\end{equation}
The value of $\tau(a)$ is plotted in Fig. \ref{fig:tau}, for our stellar model and for $M_2= 0.6~\rm{M_{\odot}}$.
We note that the tidal interaction is small compared with the gravitation drag inside the envelope.
At the radius where the released gravitational energy equals the envelope binding energy, $a_{\rm f1}$,
$\tau_{\rm f} = \tau(a_{\rm f1}) \simeq 0.95 \days$. 
{}From Fig. \ref{fig:Ebind_EG_r} it is evident that most of the gravitational energy is released very close to $a_{\rm f1}$.
For example, $90$ per cent ($80$ per cent) of the energy is released when the companion spirals-in from $14~\rm{R_{\odot}}$ 
($7.1~\rm{R_{\odot}}$) to $a_{\rm f}$, 
and the timescale in that location is $\tau(14~\rm{R_{\odot}}) \simeq 6.1 \days$ 
($\tau(7.1~\rm{R_{\odot}}) \simeq 2.4 \days$). 
Overall, most of the energy is deposited in the inner region of the envelope and during a time much shorter than the dynamical time of the outer regions.
The consequences of this instantaneous energy liberation are discussed in the following sections.

We note that had we taken the spiral wave dissipation timescale instead of the tidal timescale
(e.g., from fig. 3 in \cite{SandquistTaam1998}),
we would have gotten a shorter timescale than $\tau_{\rm{tid}}$ (the blue line in Fig. \ref{fig:tau} would
have been located a little lower).
But, according to equation (\ref{eq:tau}) the relevant timescale for $\tau$ would not have changed.

\section{THE BLAST WAVE PROPAGATION THROUGH THE AGB ENVELOPE}
\label{sec:blast}

We perform a self similar calculation of a blast wave propagating from the interior of the AGB outwards.
The calculation is based on the work of \cite{Sedov1959}.
To facilitate the analytic solution we consider two limiting cases of the short energy deposition discussed above.
In the first case we assume an instantaneous release of the binary gravitational energy (section \ref{sec:blast:instantaneous}),
while in the second case the energy is released at a constant rate over a short period (section \ref{sec:blast:continuous}).

\subsection{Instantaneous energy release}
\label{sec:blast:instantaneous}

\begin{figure}[t]
\resizebox{0.5\textwidth}{!}{\includegraphics{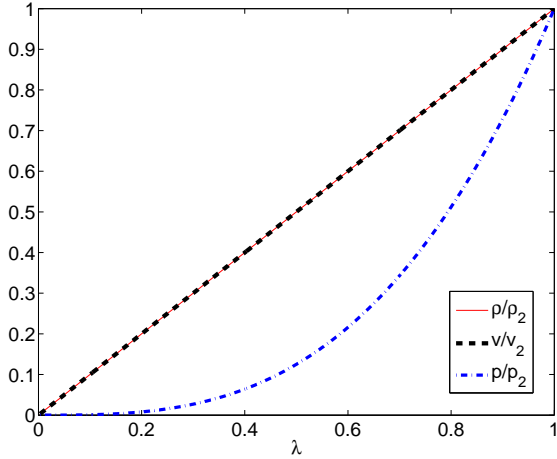}}
\caption{\footnotesize{The asymptotic self-similar solution of an intense explosion blast wave which satisfies $\omega=(7-\gamma)/(\gamma+1)$,
as solved by \cite{Sedov1959}.}}
\label{fig:Blast14_selfsimilar}
\end{figure}

Here we assume that the energy is deposited instantaneously at $t=0$.
In this case (based on chapter 14.4 in \citealt{Sedov1959}) we neglect the initial pressure in the AGB envelope and
consider an adiabatic motion of a perfect gas.
The density profile before the blast is given by equation(\ref{eq:rho}), and the envelope is static.
The general solution for the blast wave is rather complicated for a general value of $\omega$, and adiabatic exponent $\gamma$.
However when the relation $\omega=(7-\gamma)/(\gamma+1)$ holds, as in the case of the AGB model we use with $\omega=2$ and $\gamma=5/3$,
there is a simple asymptotic solution for the velocity, density, and pressure behind the shock, given respectively by
\begin{equation}
\begin{split}
v &= v_2 \lambda = \frac{2}{3\gamma-1} \frac{r}{t} \quad \quad\\
\rho &= \rho_2 \lambda = \frac{A(\gamma+1)}{r^\omega(\gamma-1)}\lambda^{\frac{8}{\gamma+1}}\\
p&= p_2 \lambda^3 =\frac{A}{r^{\omega-2}t^2} \frac{2(\gamma+1)}{(3\gamma-1)^2}\lambda^{\frac{8}{\gamma+1}},
\end{split}
\label{eq:blast_intense}
\end{equation}
where the index $2$ refers to the immediate postshock values, and where $\lambda$ is the self-similar variable
\begin{equation}
\lambda=\left(\frac{A\alpha}{E_0}\right)^{\frac{1}{5-\omega}} r t^{-\frac{2}{5-\omega}} = \frac{r}{r_2},
\label{eq:blast_intense_lambda}
\end{equation}
with $r_2$ the location of the shockwave, and $E_0$ the energy deposited in the envelope.
The constant $\alpha$ is determined from energy conservation.
The self-similar solution for $\omega=2$ and $\gamma=5/3$ is shown in Fig. \ref{fig:Blast14_selfsimilar}.

\begin{figure*}[t]
\resizebox{1\textwidth}{!}{\includegraphics{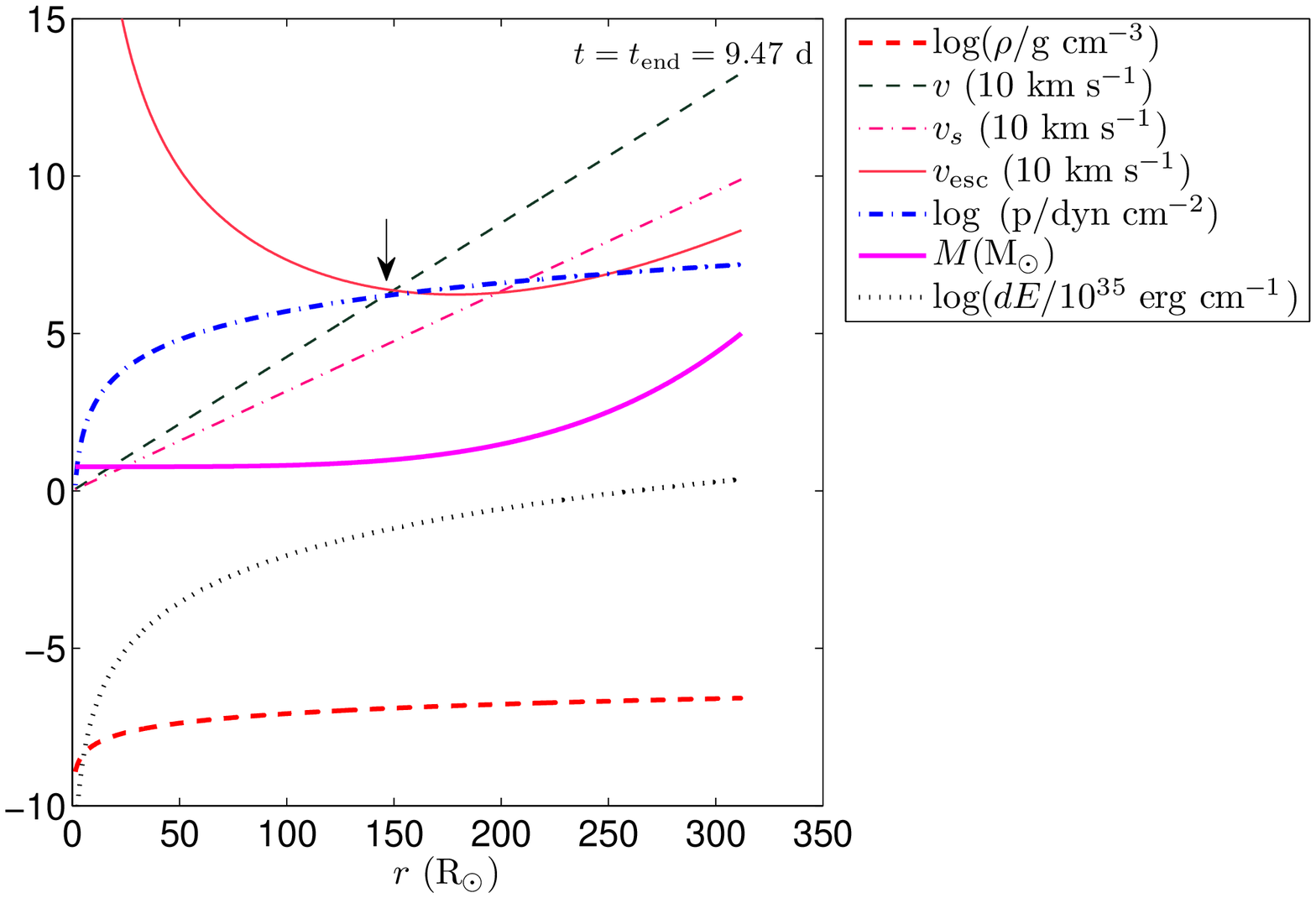}}
\resizebox{1\textwidth}{!}{\includegraphics{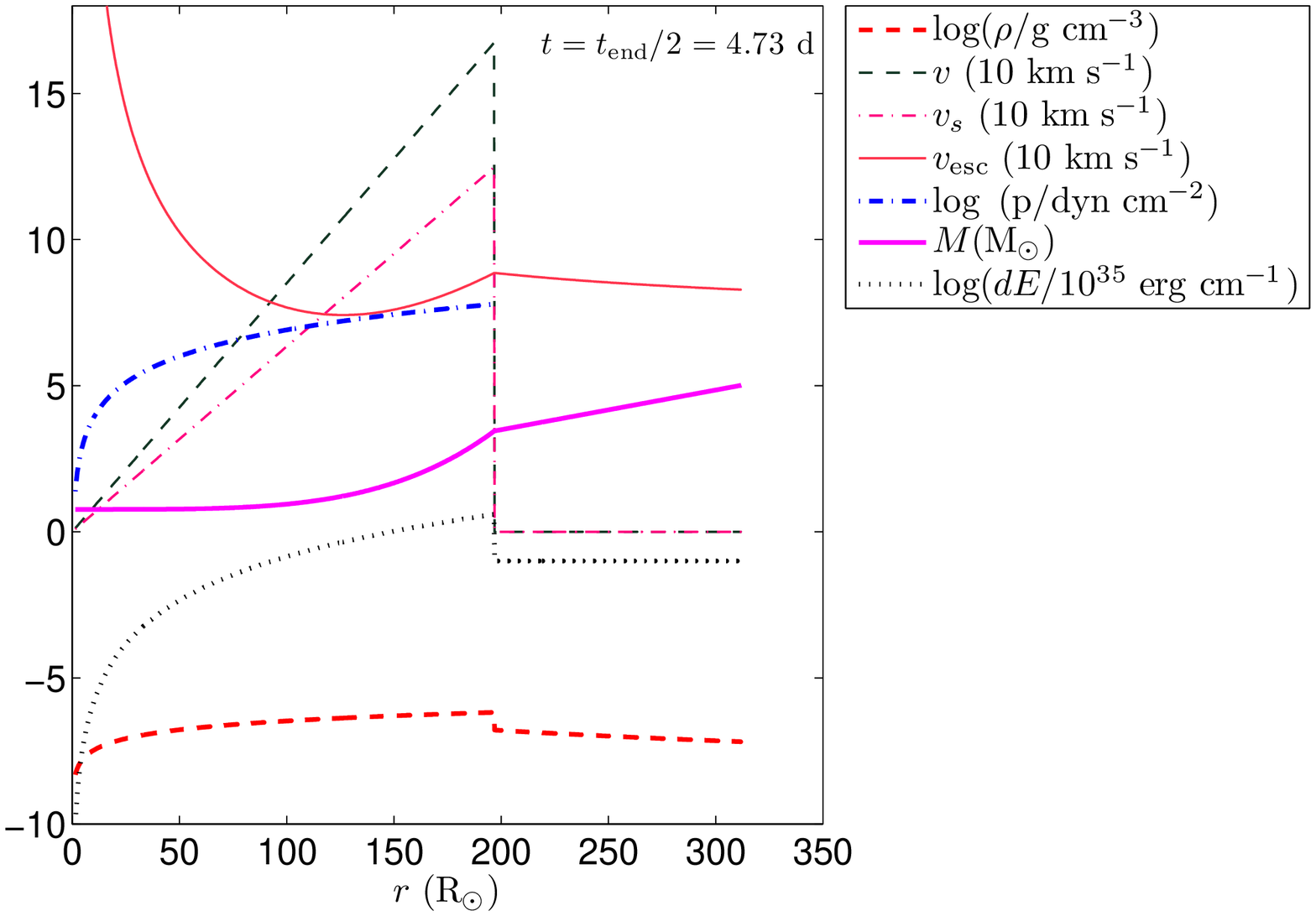}}
\caption{\footnotesize{The solution of an intense explosion blast wave, for an instantaneous release of energy inside the
AGB envelope.
We use AGB envelope mass $M_{\rm{env}}=4.2~\rm{M_{\odot}}$ and a stellar radius of $R_\star=310~\rm{R_{\odot}}$. 
The companion mass is $M_2=0.6~\rm{M_{\odot}}$.
The physical quantities are plotted at the time the reaches the stellar surface ($t_{\rm{end}} = 9.47 \days$, upper panel),
and at half of that time ($t= t_{\rm{end}}/2 = 4.73 \days$, upper panel).
The plotted physical quantities are the density of the envelope $\rho$, the velocity $v$, the sound velocity $v_{\rm s}$,
the escape velocity $v_{\rm{esc}}$, the pressure $p$, the mass $M$ inner to a radius $r$ (including the core mass), and
the total energy $dE$ in a shell $dr$.
In the upper panel we find that the intersection of the post shocked velocity $v$ with the escape velocity $v_{\rm{esc}}$ occurs
at a radius of $\sim 150~\rm{R_{\odot}}$ (marked with a small arrow). 
The mass inner to this point does not reach the escape velocity and falls back to the binary system.
We find the fall-back mass to be $M_{\rm{fb}} \simeq 0.22~\rm{M_{\odot}}$.}}
\label{fig:Blast14_tend}
\end{figure*}

In addition to the energy deposited by the spiraling-in companion, the envelope itself has thermal energy $E_{\rm{th}}$,
which we also add to that energy.
We use the virial theorem $E_{\rm{th}} = 0.5 E_{\rm bind}$ to find the energy deposited in the envelope
\begin{equation}
E_0 = E_{\rm bind}+ E_{\rm{th}} = \frac{3}{2} E_{\rm bind}
\label{eq:E0}
\end{equation}
The justification for considering the thermal energy of the envelope was recently discussed by \cite{DeMarco2011}.
We use this energy in equation (\ref{eq:blast_intense_lambda}) to obtain the solution that is shown in Fig. \ref{fig:Blast14_tend}.

We use the solution to estimate the mass that stays bound to the remnant binary system.
We calculate the escape velocity $v_{\rm{esc}}(r) = [2G(M(r)+M_2)/r]^{1/2}$
taking the companion mass into account, as its mass contributes to the gravitational force
at the time the envelope is ejected.
Radiation pressure is not an important factor in expelling material outwards (e.g., \citealt{Passy2011}).
We therefore neglect it, and set the condition for ejecting material to have its velocity larger than the escape velocity.
The amount of material that does not reach the escape velocity and hence remains bound is determined from the condition at the end
of the calculation $v(r,t_{\rm{end}})<v_{\rm{esc}}(r,t_{\rm{end}})$, where $t_{\rm{end}}$ is the time when the shockwave reaches $R_\star$.
This material is expected to fall-back towards the center, as the pressure decreases inwards (Fig. \ref{fig:Blast14_tend}).
For our case we find $t_{\rm{end}} = 9.5 \days \simeq 10 \tau_{\rm f}$. 
This shows that our assumption that the energy is released instantaneously is quite adequate here.

\begin{figure}[t]
\resizebox{0.5\textwidth}{!}{\includegraphics{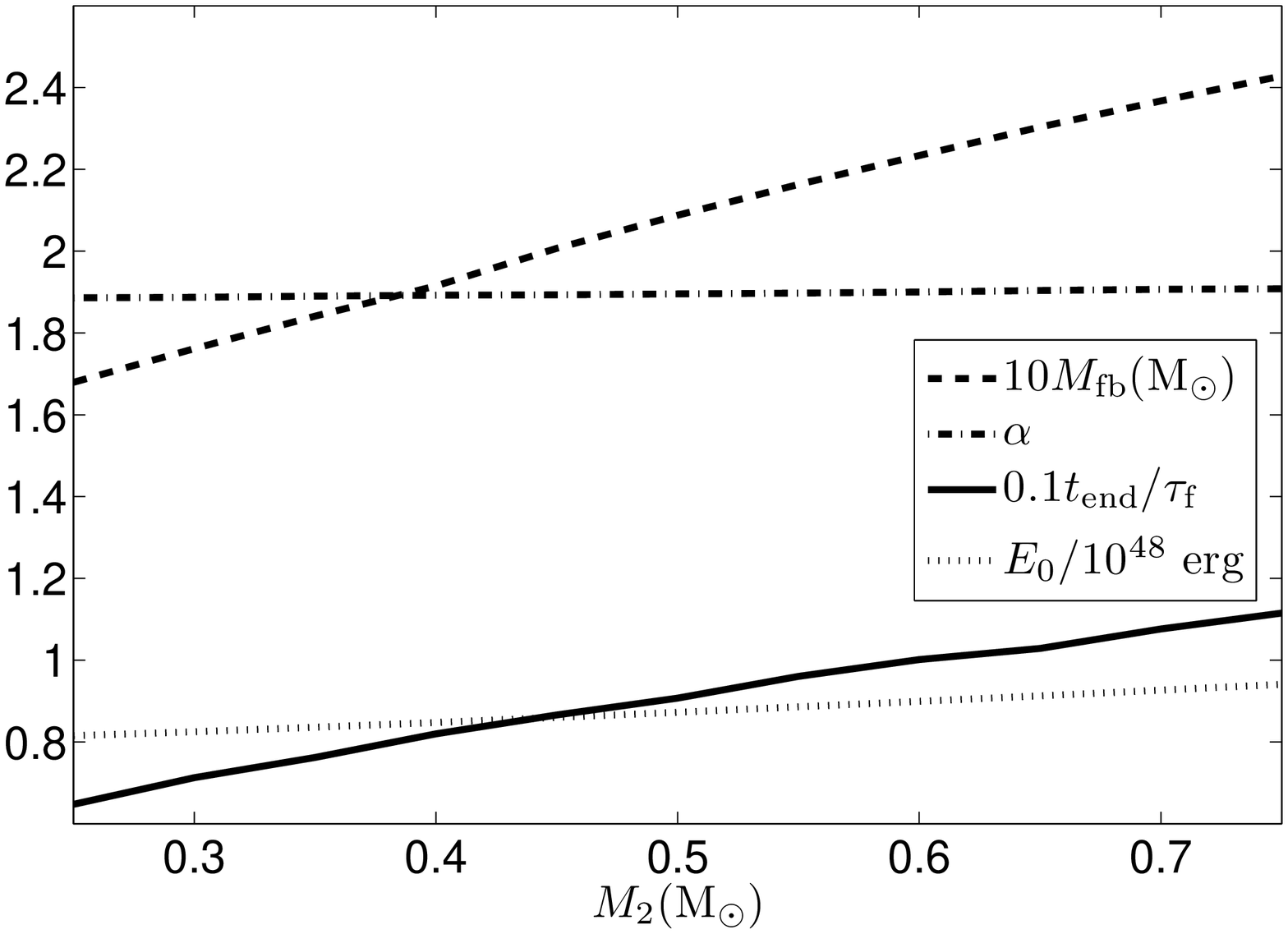}}
\caption{\footnotesize{
Physical properties as a function of the companion mass $M_2$:
the spiraling-in timescale at the final orbital separation ($\tau_{\rm f}$),
the energy deposited in the envelope $E_0$ (equation \ref{eq:E0}),
the constant $\alpha$ (equation \ref{eq:blast_intense_lambda}),
the fall-back mass $M_{\rm{fb}}$,
and the blast-wave propagation time to envelope expel time ratio $t_{\rm{end}} / \tau_{\rm f}$.}}
\label{fig:Blast14_Mfb}
\end{figure}

We find that for $M_2=0.6~\rm{M_{\odot}}$, our characteristic case, the fall-back mass is $M_{\rm{fb}} \simeq 0.22~\rm{M_{\odot}}$. 
For larger $M_2$ more of the envelope mass is falling back, as we show in Fig. \ref{fig:Blast14_Mfb}.
In Fig. \ref{fig:Blast14_Mfb} we also show the ratio between the time it takes for the blast wave to reach the AGB radius ($t_{\rm{end}}$)
and the spiraling-in timescale at the final orbital separation ($\tau_{\rm f}$).

The formation of accretion disks around one or two of the post-CE stars by fall back material in the post-AGB phase was discussed before,
but at much later times after the end of the CE phase (e.g., \citealt{Soker2001}).
Here we discuss a formation of a disk immediately as the main CE phase ends, and emphasize a disk around both stars--a circumbinary disk.

The total internal energy of the gas should include ionization energy as well,
as the gas cools and recombines when it is expelled.
Here we include only the thermal energy.
The ionization energy might become an important source in extended AGB stars that have low binding energy per
unit mass (e.g., \citealt{Han2002}; see \citealt{SokerHarpaz03} for a different view).
For AGB stars the effects of ionization energy in the internal energy of the gas may be important.
As we consider here massive AGB stars, it seems the ionization energy will not change much the
amount of expelled mass.
While the binding energy goes as $E_{\rm bind} \propto M^2$, the ionization energy goes as $E_{\rm i} \propto M$.
For an envelope of $M_{\rm{env}}=4.23~\rm{M_{\odot}}$ the ionization energy available is $\sim 10^{47} \erg$, while the energy deposited
in our calculation is $\sim 9 \times 10^{47} \erg$.
Furthermore, the bound mass resides in the inner part of the expanding envelope.
The recombination radiation of the bound mass will help in part to expel the mass further out.
Therefore, the $\sim 11$ per cent additional energy of recombination will not be efficiently deposited into
the bound mass itself.
It might expel an extra small amount of mass that is just barely bound.

In our analytic study we are limited to spherically symmetric geometry.
In asymmetrical cases we expect that more mass will stay bound. The reasoning goes as follows.
Part of the mass stays bound because a fraction of the envelope leaves with much more
energy than is required to escape, hence leaving part of the envelope with a total negative energy.
An asymmetric effect, either a denser envelope in the equator or more energy injection into the
equatorial plane, for example, will increase the uneven energy distribution in the envelope.
More energy deposition in the equatorial plane might leave more mass bound near the polar directions.
The question is what is exactly the value of the specific angular momentum of the bound envelope.
It is possible that interaction of the bound mass with the binary will lead to the formation of a
circumbinary disk even when the initial specific angular momentum is small.

\subsection{Continuous energy deposition}
\label{sec:blast:continuous}

We find that the timescale for binary energy deposition $\tau_{\rm f}$ is smaller but not negligible relatively to the time for the
shock to reach the stellar surface $t_{\rm end}$.
For that we now turn to calculate a case where the energy is deposited continuously during a finite time.
In this second case we do not neglect the pressure of the envelope
The self similar calculation of the blast wave we perform in this section
is based on chapter 6.6 of \cite{Sedov1959}.
For this self similar solution the energy deposition rate is $E_0 \propto t^{2(5-2\omega)/\omega} \propto t$,
where the last equality is for $\omega=2$.

As a result of the continues energy deposition and the counter pressure of the pre-shock envelope,
a void is created in the inner region, and the post-shock material is contained within the blast shell,
defined as the volume between the outer boundary of the void and the shockwave.
The blast shell moves outwards, until it reaches the stellar surface.
The problem with this description is that in our case the energy source is the companion.
When the inner boundary of blast shell crosses the companion, there is no mechanism to transfer
the energy from the central engine to the shell, because the companion is inside the void.
We therefore limit ourselves for treating a blast with a continuous energy release to low
radii and short times, when the blast shell is close to the center.
As in our case the continuous energy release solution is relevant only to short times, we stop the energy repositioning at $\tau_{\rm f}$.

The self similar variable here is
\begin{equation}
\lambda= r (\beta A G)^{\frac{-1}{\omega}} t^{\frac{-2}{\omega}} =\frac{r}{r_2}.
\label{eq:continuous_selfsimilar}
\end{equation}
The self-similar functions are $R$, $V$, $Z$ and $\mathcal{M}$, defined as (chapter 6.6 in \citealt{Sedov1959})
\begin{equation}
\begin{split}
\rho(r,t)&=\frac{1}{Gt^2}R(\lambda)
\\
v(r,t)&=\frac{r}{t}V(\lambda)
\\
p(r,t)&=\frac{r^2}{Gt^4}P(\lambda)=\frac{r^2}{\gamma Gt^4} R(\lambda) Z(\lambda)
\\
M(r,t)&= \frac{r^3}{Gt^2} \mathcal{M}(\lambda)
\end{split}
\label{eq:continuous_selfsimilar_variables}
\end{equation}
The differential equations that describe the blast, in terms of the self similar functions are
\begin{equation}
\begin{split}
\frac{{dR}}{d \lambda} &= -\frac{R\, \left[\mathcal{M} - V - \left(3\, V - 2\right)\, \left(V - \frac{2}{\omega}\right) + \frac{2\, Z}{\gamma} + V^2 - \frac{2\, Z\, \left(V + \gamma - 2\right)}{\left(V - \frac{2}{\omega}\right)\, \gamma}\right]}{\lambda\, \left[Z - {\left(V - \frac{2}{\omega}\right)}^2\right]}
\\
\frac{{dV}}{d \lambda} &= \frac{\left(V - \frac{2}{\omega}\right)\, \left[\mathcal{M} - V - \left(3\, V - 2\right)\, \left(V - \frac{2}{\omega}\right) + \frac{2\, Z}{\gamma} + V^2 - \frac{2\, Z\, \left(V + \gamma - 2\right)}{\left(V - \frac{2}{\omega}\right)\, \gamma}\right]}{\lambda\, \left[Z - {\left(V - \frac{2}{\omega}\right)}^2\right]} - \frac{3\, V - 2}{\lambda}
\\
\frac{{dZ}}{d \lambda} &= - \frac{\left(\gamma - 1\right)\, \left[\mathcal{M} - V - \left(3\, V - 2\right)\, \left(V - \frac{2}{\omega}\right) + \frac{2\, Z}{\gamma} + V^2 - \frac{2\, Z\, \left(V + \gamma - 2\right)}{\left(V - \frac{2}{\omega}\right)\, \gamma}\right]}{\lambda\, \left[Z - {\left(V - \frac{2}{\omega}\right)}^2\right]} - \frac{Z\, \left(2\, V + 2\, \gamma - 4\right)}{\lambda\, \left(V - \frac{2}{\omega}\right)}
\\
\frac{{d\mathcal{M}}}{d \lambda} &= -\frac{3\, \mathcal{M} - 4\, \pi\, R}{\lambda} ,
\end{split}
\label{eq:continuous_selfsimilare_quations}
\end{equation}
and the boundary conditions at the shock front are
\begin{equation}
\begin{split}
R_2 &= -\frac{2\, q\, \left({\gamma} + 1\right)\, \left(\omega - 1\right)\, \left(\omega - 3\right)}{\pi\, {\omega}^2\, {\gamma}\, \left(2\, q + {\gamma} - 1\right)}
\\
V_2 &= -\frac{4\, \left(q - 1\right)}{\omega\, \left({\gamma} + 1\right)}
\\
Z_2 &= -\frac{4\, \left(q\, \left({\gamma} - 1\right) - 2\, {\gamma}\right)\, \left(2\, q + {\gamma} - 1\right)}{{\omega}^2\, {\left({\gamma} + 1\right)}^2}
\\
\mathcal{M}_2 &= \frac{8\, q\, \left(\omega - 1\right)}{{\omega}^2\, {\gamma}}
\end{split}
\label{eq:continuous_selfsimilare_boundary}
\end{equation}
where
\begin{equation}
q = -\frac{\pi\, {\omega}^2\, {\gamma}}{2\, \left(\omega - 1\right)\, \left(\omega - 3\right)\, {\beta}},
\label{eq:continuous_q}
\end{equation}
and $\beta$ is a constant determined from energy conservation.

\begin{figure*}[t]
\resizebox{1\textwidth}{!}{\includegraphics{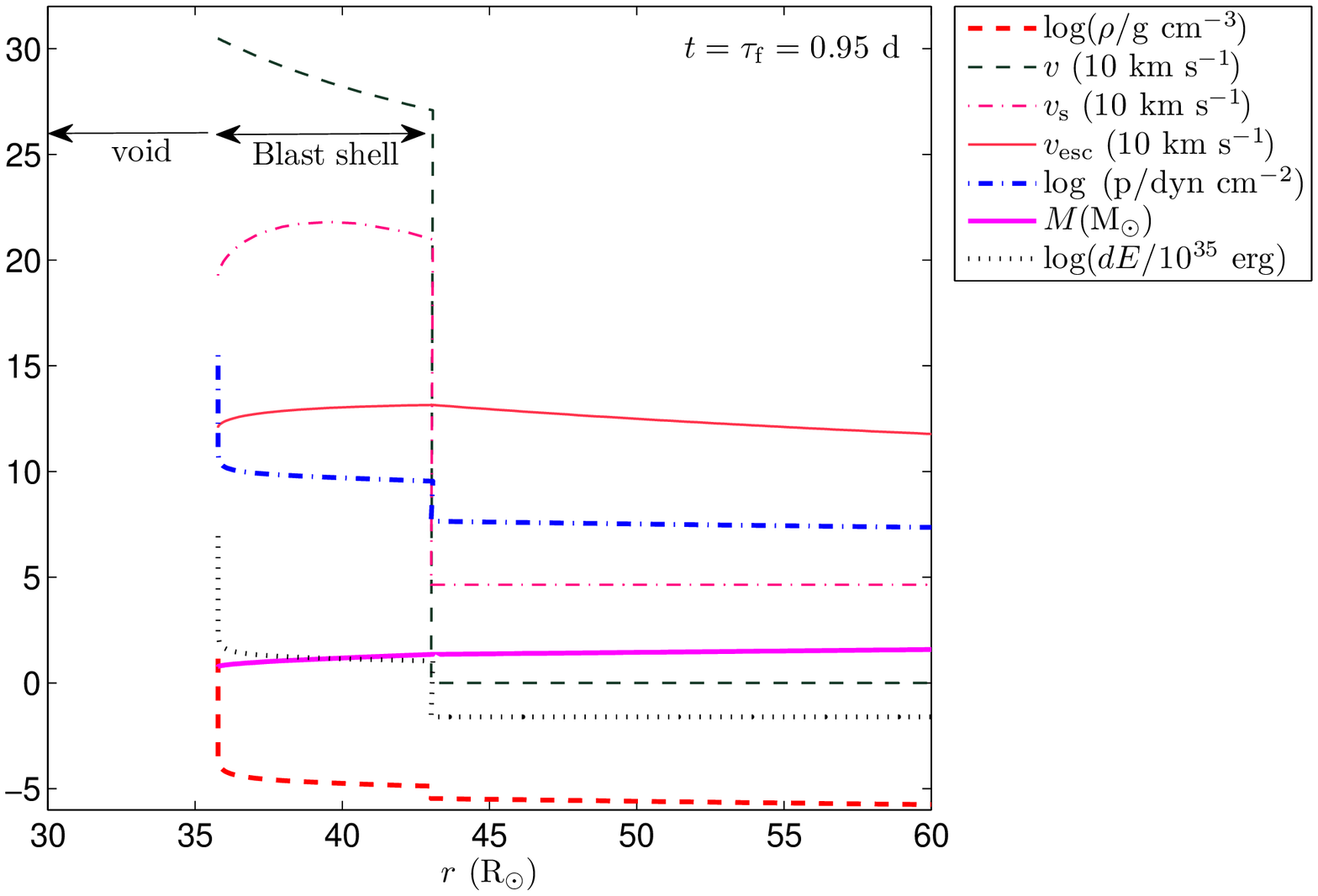}}
\caption{\footnotesize{The solution of a blast wave with continuous energy release from $t=0$ to $t=\tau_{\rm f} \simeq 0.95 \days$.
A blast shell is formed, starting from the end of the void and ending at the shockwave.
For explanations to the legend see the caption in Fig. \ref{fig:Blast14_tend}.
}}
\label{fig:Blast16_tend_invest_energy}
\end{figure*}

The solution is plotted in Fig. \ref{fig:Blast16_tend_invest_energy}.
We can see that the blast shell resides in the inner part of the envelope, and supersonically moves outward.
The energy in the blast wave is the same as that in the previous case (equation \ref{eq:E0}), such that we are back
to the intense explosion case (section \ref{sec:blast:instantaneous}).
Namely, if the process does not last for a long time (see below), the two cases studied in this section lead to the same result
that a non-negligible mass stays bound.

\section{POSSIBLE IMPLICATIONS}
\label{sec:implications}

\subsection{The role of a circumbinary disk}
\label{sec:implications:disk}

The fall-back mass is expected to posses angular momentum, originally coming from the orbital angular momentum of the companion.
We suggest that the fall-back mass forms a disk around the remnant binary system -- the companion and the leftover core of the AGB.
Even if the specific angular momentum of the fall back mass is not sufficient to form a circumbinary disk, later interaction
of this gas with the binary system might lead to the formation of a disk.
Such an interaction is thought to lead to the formation of the circumbinary disks around post-AGB binary stars with an orbital separation of
$\sim 1 \AU$ (\citealt{vanWinckel2000}; \citealt{vanWinckel2006}; \citealt{vanWinckel2009}).
\cite{vanWinckel2009} conclude that circumbinary disc can tremendously affect the evolution of the binary system in the center,
and that the formation of circumbinary gravitationally bound disks occur in a wide range of post-AGB binaries.
In these binaries the typical orbital separation is $\sim 1 \AU$, many of them have substantial eccentricity,
and even if they avoided a CE phase, the companion had a strong interaction with the AGB envelope.

In the simulations of \cite{Passy2011} even larger larger than found in our calculations amount of material remains bound to the binary system.
However, they find that the orbital separations of the post-spiral-in phase are larger than the observed ones.
Though that \cite{Passy2011} found that there is enough orbital energy to unbind the envelope,
the binary system in their simulations remains in a completely evacuated volume and the leftover bound envelope is at
 $\sim 100 ~\rm{R_{\odot}}$.

We therefore assume that the fall back gas forms a circumbinary disk.
The binary system transfers angular momentum to the circumbinary disk, and consequently the radius of the
disk expands, the binary separation decreases and the eccentricity increases
(e.g., \citealt{Artymowicz1991}).
We assume that the structure of the circumbinary disk formed in our scenario resembles the
structure of the disk studied in \cite{Artymowicz1991}.
In the calculations of \cite{Artymowicz1991} the disk extends from the nearest stable circumbinary orbit of $\sim 2.5 a$
up to $6a$, where $a$ is the binary separation.
We substitute for the binary system used in section \ref{sec:blast},
and take the binary system to be composed of $M_{\rm{core}}=0.77~\rm{M_{\odot}}$ and $M_2=0.6~\rm{M_{\odot}}$ components
at a separation of $a_{\rm f} \simeq 1.4~\rm{R_{\odot}}$, which is the final binary separation
of the CE phase.
We assume that the initial binary orbit is circular ($e_0=0$).
The circumbinary disk mass is the most uncertain quantity here.
We here scale it with $M_{\rm{disk,0}} = M_{\rm{fb}} \simeq 0.2~\rm{M_{\odot}}$.
Scaling the results of \cite{Artymowicz1991} by these values we find that the rate of change
of the semi-major axis is
\begin{equation}
\frac{\dot{a}}{a} \simeq -3 \times 10^{-8} \left(\frac{q_d}{0.16}\right) \left(\frac{\Omega_b}{4.2 \times 10^{-4} \s^{-1}}\right) \s^{-1}, 
\label{eq:adotoveraArty}
\end{equation}
and the rate of change of the eccentricity is
\begin{equation}
\dot{e} \simeq 1.3 \times 10^{-7} \left(\frac{q_d}{0.16}\right)
\left(\frac{\Omega_b}{4.2 \times 10^{-4} \s^{-1}}\right) \s^{-1}, 
\label{eq:edotArty}
\end{equation}
where
\begin{equation}
\Omega_b=\sqrt{\frac{G(M_{\rm{core}}+M_2)}{a_{\rm f}^3}} \simeq 4.2 \times 10^{-4} \s^{-1} 
\label{eq:Omegab}
\end{equation}
is the binary orbital angular velocity, and
\begin{equation}
q_d=\frac{M_{\rm{disk}}}{M_{\rm{core}}+M_2} 
\label{eq:qd}
\end{equation}
is the disk to binary mass ratio.
At the end of the CE, when the circumbinary disk --binary interaction starts,
the disk to binary mass ratio is $q_{d,0} \simeq 0.16$.

Another effect we take into account is the dissipation of the circumbinary disk.
The energy released by the shrinking binary orbit is transferred to the disk.
Because the total power is larger than the Eddington limit, a large fraction of the energy
is carried by gas escaping from the disk.
We take this mass loss into account by equating the energy carried by the wind to that
released by the shrinking binary system.
As material probably escape from the entire disk, we take as an approximation the
wind to leave from the middle of the disk, $R_{\rm wind}(t) = 4.25 a(t)$.
More likely, more mass will leave from the inner region of the disk,
hence less mass can carry the same amount of energy.
This will leave more mass in the disk, making the scenario discussed here more efficient even.
In general, winds escape from stars and disks with a velocity about equal to the escape velocity.
Therefore, the total energy carried by the wind is the sum of the binding energy of the gas
and the kinetic energy at infinity.
We consider also the kinetic energy of the wind, assuming its terminal velocity is the
escape velocity from $R_{\rm wind}$.
The kinetic energy introduces a factors of $\chi$ in equation (\ref{eq:dMdisk}) below.

We use equations (\ref{eq:adotoveraArty}) and (\ref{eq:edotArty}) to integrate the semi-major axis
decrease and eccentricity increase due to the circumbinary disk -- binary interaction.
We take the circumbinary disk mass to reduce according to the released
orbital energy from the reduction of the binary orbit
\begin{equation}
dE(t)= - \frac{G M_{\rm{core}} M_2}{2a_{\rm f}} + \frac{G M_{\rm{core}} M_2}{2a(t)}.
\label{eq:dE}
\end{equation}
The binding energy of the escaping disk material per unit mass is
\begin{equation}
\frac{dE_{\rm{bind,disk}}(t)}{dm}= - \frac{G(M_{\rm{core}}+M_2)}{2R_{\rm{wind}}},
\label{eq:Ebinddisk}
\end{equation}
and the kinetic energy is parameterized as
\begin{equation}
\frac{dE_{\rm{kin,disk}}}{dm}=(\chi-1)\frac{dE_{\rm{bind,disk}}}{dm}.
\label{eq:Ekindisk}
\end{equation}
Hence, the rate of change of the disk mass is
\begin{equation}
dM_{\rm{disk}}(t)= - \frac{dE(t)}{\chi\frac{dE_{\rm{bind,disk}}(t)}{dm}}.
\label{eq:dMdisk}
\end{equation}

Our results for an initial eccentricity $e_0=0$ are presented in Fig. \ref{fig:AE}.
The decrease in orbital separation and increase in eccentricity are clearly seen.
The circumbinary disk -- binary interaction terminates when one of two things occur:
either the circumbinary disk is completely depleted, or the two stars merge (occurs when $ e \simeq 1$).
We term the time period of the circumbinary disk -- binary interaction $\tau_1$.
As can be understood from Fig. \ref{fig:AE}, taking $e_0>0$ would shorten $\tau_1$,
so taking $e_0=0$ is a conservative parameter selection.
As stated above, the most uncertain parameter is the initial disk mass,
which sets the value of $q_{d,0}$ (equation \ref{eq:qd}).
We repeat the calculation for a few values of $q_{d,0}$.
For the case where the disk wind is ejected at the escape velocity (namely $\chi=2$)
we find that the circumbinary disk -- binary interaction leads to a merger for $q_{d,0} \ga 0.12$,
without the necessity of a following process (emission of gravitational waves (GW),
discussed in section \ref{sec:implications:coredegenerate} below).

We note that post-CE circumbinary disks were suggested under other assumptions about the CE evolution.
The formation of a post-CE thick circumbinary disk was invoked by \cite{Soker1992} in a case of
a more gradual spiraling-in process where part of the envelope remains in an hydrostatic equilibrium
rather than falling back.
A differentially rotating thick disk or torus was found in intermediate stages of the CE evolution in the
3D numerical simulations of \cite{SandquistTaam1998}.
In these cases a rapid merging is expected as well.

\cite{Ivanova2011} suggests another mechanism for post-CE merger, applicable for stars with non-degenerate cores.
\cite{Ivanova2011} divides the CE phase into two phases.
In the first stage the envelope is expelled, and a remnant is left in the center.
This stage takes $\sim 1 \yr$.
Some envelope mass is left above the core.
In the second stage the left-over envelope undergoes thermal readjustment, and its outer region expands.
This occurs on a timescale of a few $\times 10^3 \yr$.
\cite{Ivanova2011} concludes that in many cases the left-over envelope would expand beyond the companion,
forming another CE phase that substantially increase the merger probability.

As we found above that some of the material in the envelope falls back towards the binary system,
we suggest that the readjustment phase in the case studied here can be even more dramatic than
in the case studied by suggested by \cite{Ivanova2011}.
In the case studied here there are two possible scenarios.
In the first scenario the falling back material will be added to the
remnant material and will be included in the readjustment phase.
This material will be added to the outer layer of the left over envelope, which has the potential to expand and swallow the companion.
In the second scenario, where there is enough angular momentum, a circumbinary disk is formed and the binary separation
decreases as discussed above.
Overall, our new finding of the existence of a large amount of fallback material supports the conclusion of \cite{Ivanova2011} of a higher percentage of immediate post-CE mergers, although \cite{Ivanova2011} mechanism works for giants with non-degenerate core.

We emphasize the role played by the circumbinary disk.
At early stages of the CE the spiraling in is relatively slow and the rate of energy release by the spiraling-in
companion is low.
At that stage the convection in the AGB envelope efficiently transfers the energy outwards.
Most (but not all) of the extra energy is radiated away at this early phase.
In the final phase of the CE, when the companion is deep inside, energy released at a very high rate, under the assumption
used here.
In that case energy is carried by the expelled envelope.
This stops the spiraling-in process.
Here we suggest that further spiraling-in can take place due to interaction with a circumbinary disk.
The circumbinary disk is an efficient mechanism to absorb the binary orbital angular momentum.
In addition, the circumbinary disk dissipates the binary orbital energy
and radiates it by expelling mass and radiation from its surface.
This phase ends either with a merger or with a very small orbital separation.
A final small orbital separation will be interpreted as a small $\alpha_{\rm CE}$.

The value of $\alpha_{CE}$ will be small even if our assumption of a rapid release of energy at the end of the CE is not satisfied,
but rather the release of energy is a long process (e.g., $\sim 1\yr$, \citealt{SandquistTaam1998}; $\sim 10\yr$, \citealt{DeMarco2003}).
\cite{DeMarco2011} suggest that if the spiral-in of the companion takes longer than the dynamical time,
as in that case of long time energy release, the
thermal energy of the AGB will be used to help unbind the envelope.
Therefore, they conclude it is expected that the companion will spiral-in inward to the radius predicted by taking $\alpha_{\rm CE}=1$

\subsection{The SN Ia scenario from an early merger}
\label{sec:implications:earlymerger}

Here we discuss some possible outcomes of WD-WD (double degenerate) merger,
where the two component are made of CO.
The WD-WD merger is one of the scenarios that lead to SNe Ia (\citealt{IbenTutukov1984}; \citealt{Webbink1984}).
Most simulations and calculations of double degenerate merger (e.g., \citealt{Yoon2007}) assume the merger to
occur a long time after the CE phase, when the two WDs are already cold.
In the scenario we propose here, the merger occurs within the final stages of the CE, while the core is still hot.
Below we speculate that some of these will become SNe Ia after a very long time dictated by angular momentum loss.
The speculation is based on studies of double degenerate mergers that we describe below.
We then discuss their application to the scenario proposed here.

In analyzing the outcome of the double degenerate merger we must take into account the following.
(i) In the merger process the more massive WD remains at the center of the merger product, while the lighter WD
is destructed and becomes the envelope, and possibly a thick disk around the center.
(ii) The post-AGB core is hot as it had no time to cool, while the companion WD is colder as it had the time to
cool since its formation.

In general, the merger of two WDs can end in one of three basic products (e.g., \citealt{YoonLanger2005}):
SN Ia, core collapse to a neutron star (NS), or a more massive WD.
In the first two cases, where the mass of the merger is above the critical limit,
the most important factor in determining whether there will be an explosion as a SN Ia or not
is the nature of the carbon ignition.
In the third case, the mass of the merger is below the critical limit.
We now discuss the three cases.

(1) In the first case, the companion mass is larger than the post-AGB core mass, and therefore the companion will become
the cold core of the merger (assuming it had enough time to cool) and the destructed hot post-AGB core will be its envelope.
We note that before the merger but already in the CE phase, the center of mass will be closer to the companion, and
the envelope will arrange itself as if the companion were the giant core.
In this case carbon ignition is more likely to occur during the merger processes and be off-center (\citealt{SaioNomoto1985}).
Most of the WD merger models indicate that a non-violent off-center ignition that occurs during the merger process,
burns carbon and oxygen into oxygen, neon, and magnesium.
As not enough energy remains to unbind the WD when the mass reaches the critical limit it will gravitationally collapse
and form a NS rather than SN Ia (\citealt{NomotoIben1985}; \citealt{SaioNomoto1985}; \citealt{MochkovitchLivio1990};
\citealt{HillebrandtNiemeyer2000}; \citealt{YoonLanger2005}).
In that case a NS will be formed (e.g., \citealt{YoonLanger2005}).
It is interesting to mention the possibility that in such a scenario planets
can form around the NS from SN fallback material (\citealt{Hansen2009}).

However, in many cases no carbon ignition will occur during the merger process \citep{Yoon2007}, and if the mass reaches the critical mass
a SN Ia might occur after the WD slows down.

(2) In the second case, the companion mass is smaller than the post-AGB core mass, and therefore the center
of the product will be hot.

\cite{Yoon2007} raised the possibility that a hot remnant core is less likely to ignite carbon off-axis (in the envelope) during WD-WD merger.
The reason is that a hot WD is larger, such that its potential well is shallower
and the peak temperature of the accreted destructed-WD material is lower.
Hence, in such a case supercritical-mass remnant is more likely to ignite carbon in the center at a later time, leading to a SN Ia.
The merger remnant becomes a rapidly rotating massive WD, that can collapse only after it loses sufficient angular momentum.
\cite{Yoon2007} parameterized the angular momentum loss, and found that explosion can occurs as late as $\sim 10^6 \yr$ after the merging occurs.

(3) The Third possibility is that the merger is below the critical mass.
In this case the merger will remain a WD that will continue to cool, and neither a SN Ia nor collapse are expected.

However, there is also an exceptional case.
\cite{vanKerkwijk2010} suggest that in cases where the two CO WDs have an approximately equal mass below the critical limit,
the WD merger will be fully mixed and hottest in its center.
Initially, the merger rotates differentially, but in later times some of the mass is expelled,
and a uniformly rotating core near the mass-shedding limit is obtained.
The structure is of a core which contains $\sim 80$ per-cent of the mass,
surrounded by a sub-Keplerian, very dense, partially degeneracy-pressure supported disk.
In the model of \cite{vanKerkwijk2010} the disk accretes at a high rate that leads to compressional heating that is likely
to cause central carbon ignition.
This ignition occurs at densities for which pure detonations lead to events similar to SNe Ia (\citealt{vanKerkwijk2010}).
If such a scenario occurs in our case, then the explosion will be surrounded by the expelled common envelope.

\subsection{Core-degenerate merger at the termination of the common envelope}
\label{sec:implications:coredegenerate}

Our new suggestion is that there are many double degenerate merger events where one of the components, the post-AGB core, is hot.
The timescale for angular momentum loss might be very long ($> 10^6 \yr$ after merger) even when the remnant
rotation is modest: not fast enough to slow down rapidly, but fast enough to influence the critical mass.
We hence suggest that some SNe Ia might originate from a double degenerate merger
that occurs at the last phase of the CE or shortly afterwards during the planetary nebula phase, rather than a long time after.
We term this scenario core-degenerate (CD), as one of the WD has just emerged from the core of the AGB star.

The immediate implication of processes that reduce the orbital separation at the final stages of the CE (and later form an early merger)
is that the effective value of $\alpha_{\rm CE}$ is considerably smaller than 1.
The reduction in the orbital separation can lead to merger immediately after the CE phase, or even before the entire envelope
has been lost.
In a some cases the final orbital separation will be very small, such that the merger will occur after the CE phase,
but while the core is still hot.
We now examine this case.

As mentioned in section \ref{sec:implications:earlymerger}, \cite{Yoon2007} showed that if the merger has a hot remnant core it is less likely to
ignite carbon off-center, and therefore a massive CO WD is formed.
The reason is as follows.
If the post-AGB core is still hot when it mergers with a lighter WD companion, the core radius is still
relatively large (e.g., \citealt{KoesterSchoenberner1986}; \citealt{Bloecker1995}).
When the post-AGB core radius is larger, then the gravitational potential well on the surface, where the companion mass is accumulated, is smaller.
Therefore, when the companion WD merges with the post-AGB core, the peak temperature of the merger will be in an outer radius, and in a lower temperature.
So the hotter the post-AGB core, the colder the merger peak temperature will be, for a given core mass.
An increase by a factor of 1.2 in the post-AGB core radius results in a factor of $\sim 1.2$ decrease in the mergers peak temperature.
Such a decrease might be enough to prevent off-center carbon ignition and the collapse of the merger (\citealt{Yoon2007}).
Instead, the merger product will later produce a SN Ia.

We consider the condition for a core-degenerate merger while the core is still $\sim 1.2$ larger than its final radius.
As a post-post-AGB core cools its radius decreases to an asymptotic value of a cold WD.
More massive cores cool faster (e.g., \citealt{Bloecker1995}).
For a SN Ia progenitor we take a post-AGB core of $M_{\rm{core}} \sim 0.7$--$0.8~\rm{M_{\odot}}$ with a WD companion mass
of $M_2 < M_{\rm{core}} \sim 0.6$--$0.7~\rm{M_{\odot}}$.
Such cores shrink to a radius of $1.2$ times larger than the final radius in $\tau_{\rm cool} \sim 10^5 \yr$ \citep{Bloecker1995}.
If by that time the companion WD will merge with the post-AGB core then it will be less likely to ignite carbon off-center during the merging process.

There are two possible mechanisms to further reduce the orbital separation in cases where
the circumbinary disk did not reduce the separation down to merger.
These are emission of gravitational waves and energy dissipation by tidal interaction.
The timescale for the energy dissipation by tidal interaction is much larger, even longer than the Hubble time (e.g., \citealt{Willems2010}).
We therefore consider emission of gravitational waves to be the dominant mechanism.

In an eccentric orbit WD-WD binary, the reduction of the semi-major axis due to emission of gravitational waves is faster,
as the eccentricity increases (\citealt{PetersMathews1963}).
The semi-major axis reduction and the eccentricity increase are (\citealt{PetersMathews1963}; see also equations (2) and (3) in \citealt{Ignatiev2001})
\begin{equation}
\dot{a} = -\frac{64}{5} \frac{G^3 M_{\rm{core}} M_2 (M_{\rm{core}}+M_2)}{c^5 a^3 (1-e^2)^{7/2}} \left(1+\frac{73}{24}e^2 +\frac{37}{96}e^4\right),
\label{eq:a_GW}
\end{equation}
and
\begin{equation}
\dot{e} = \frac{304}{15} \frac{G^3 M_{\rm{core}} M_2 (M_{\rm{core}}+M_2) e}{c^5 a^4 (1-e^2)^{5/2}} \left(1+\frac{121}{304}e^2\right),
\label{eq:e_GW}
\end{equation}
where $c$ is the speed of light.

\begin{figure}[t]
\resizebox{0.7\textwidth}{!}{\includegraphics{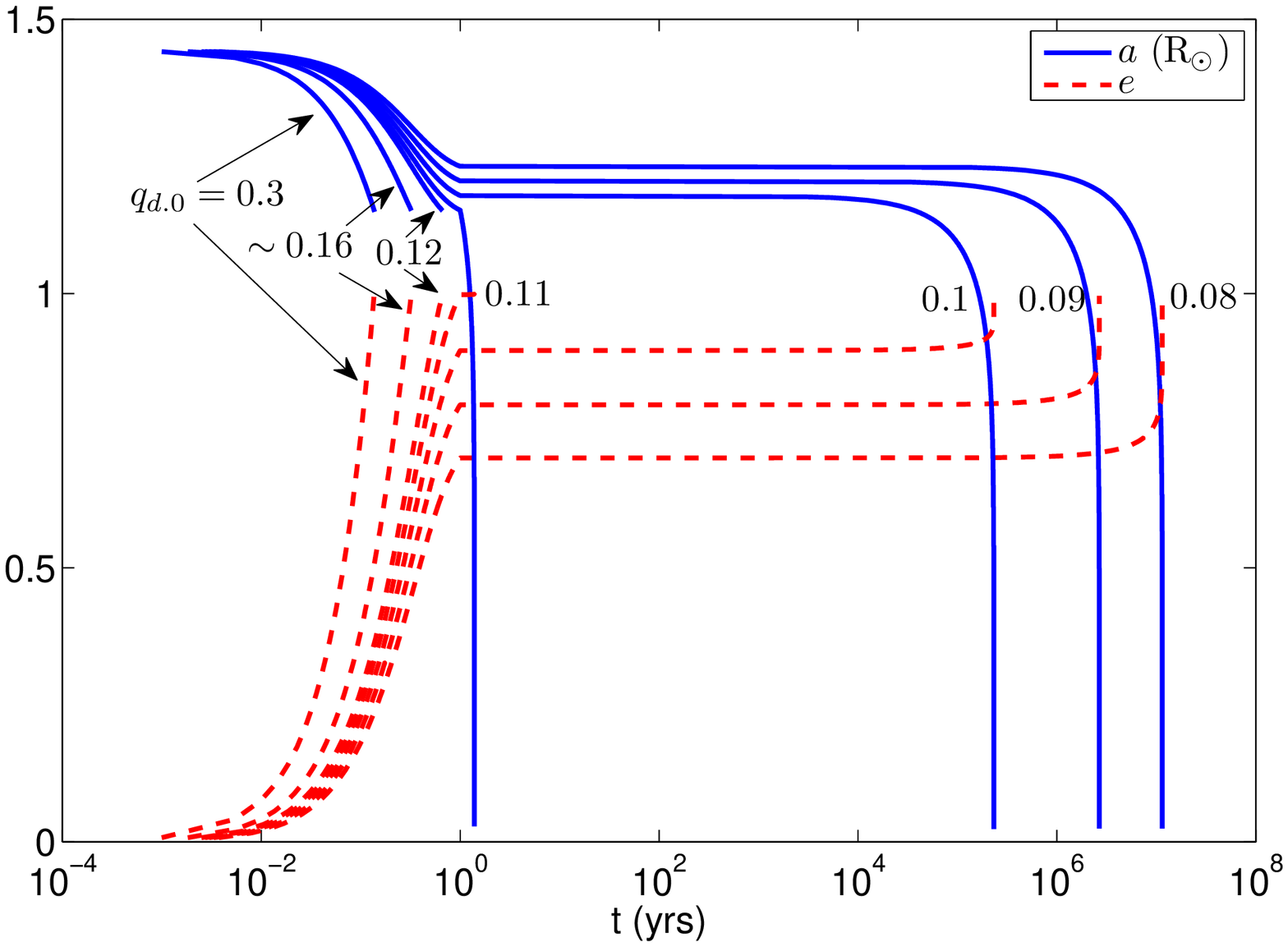}}
\caption{\footnotesize{
The variation of the semi-major axis $a$ (in solar units) and eccentricity $e$ as a function of time,
in the core-degenerate scenario.
We take $\chi=1$ in equation (\ref{eq:Ekindisk}), and an initial eccentricity $e_0=0$.
The variation occurs in two phases.
In the first phase the circumbinary disk -- binary interaction reduces $a$ and causes an increase in $e$.
The final values of $a$ and $e$ in this phase are the initial values in the second phase --
semi-major axis reduction and eccentricity increase due to emission of gravitational waves.
This phase lasts a time interval $\tau_2$, which strongly depends on the value of $e$
at the beginning of this phase, and hence on $\tau_1$.
The first phase terminates when one of two things occur: the circumbinary disk is completely depleted,
or when the binary system merge (occurs when $e \simeq 1$).
We find that the first phase is enough to create a merger for $q_{d,0} \ga 0.12$.
If the first phase has not resulted in a merger (because the circumbinary disk
is not massive enough to result in a strong interaction with the binary),
then the second phase of GW emission can reduce $a$ to the extent of merging.
We find that for $0.1 \lesssim q_{d,0} \lesssim 0.12$ the second phase will result in a merger within
$\tau_2 < \tau_{\rm cool} \simeq 10^5 \yr$.
According to the core-degenerate scenario, such an early merger that will explode as a SN Ia.
For $q_d \lesssim 0.1$ a late merger is formed, and such a merger is not expected to explode as SNe Ia.
}}
\label{fig:AE}
\end{figure}

For the core-degenerate scenario to occur, the semi-major axis reduction due to emission
of gravitational waves should result in a
merger, within a period $\tau_2 < \tau_{\rm cool}$.
The value of $\tau_2$ ,which satisfies $\tau_2 >> \tau_1$ (where $\tau_1$ is the time period of the circumbinary disk -- binary interaction)
is very sensitive to the value of $e$ from which the GW emission process starts.
This value of $e$ is determined by the final eccentricity in the circumbinary disk -- binary interaction process,
which depends on $\tau_1$.
Namely $\tau_2$ strongly depends on $\tau_1$.
Our results are presented in Fig. \ref{fig:AE}.
\begin{figure}[t]
\resizebox{0.7\textwidth}{!}{\includegraphics{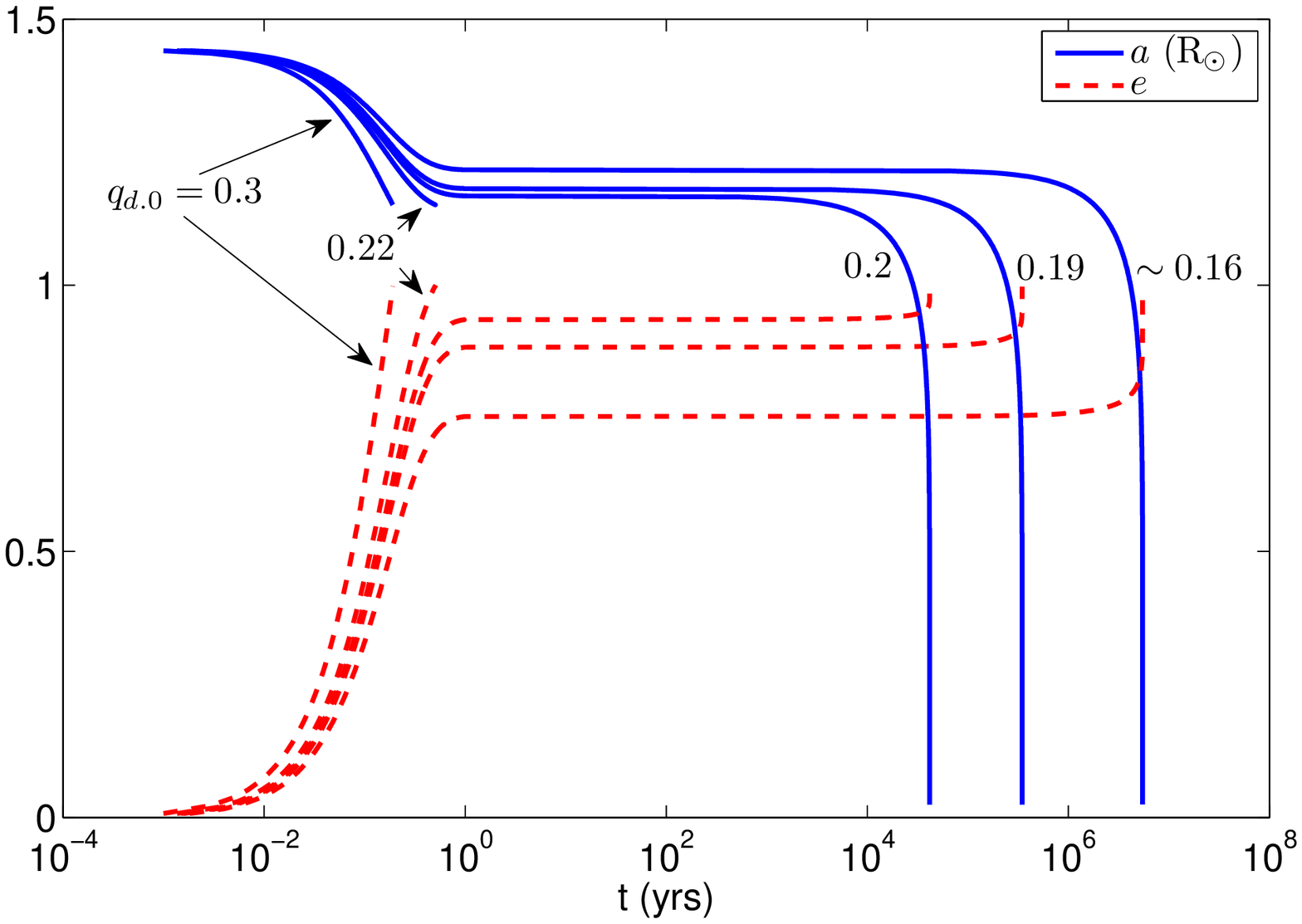}}
\caption{\footnotesize{
The same as Fig. \ref{fig:AE}, but taking $\chi=1$ in equation (\ref{eq:Ekindisk}).
Namely, with the disk wind having zero trminal velocity.
In this case we find that higher circumbinary disk mass (larger $q_{d,0}$) is required for the binary stars
to undergo a fast merge.
}}
\label{fig:AE2}
\end{figure}

We repeated our calculation (with $\chi=2$; eqution \ref{eq:Ebinddisk}) for different values of $q_{d,0}$,
and found that for $q_{d,0}\ga 0.1$
the two components will merge within a short time after the CE (Fig. \ref{fig:AE}).
As we take disk wind into account, we get that the
process of emission of gravitational waves is required for merging to occur for $0.1 \lesssim q_{d,0} \lesssim 0.12$,
but due to the interaction with the disk it will occur in a considerably shorter time
than it would have taken without it.
This shorter time can fulfil the requirement that when the core-degenerate scenario will occur,
(namely that the companion WD will merge with the post-AGB core while it is still hot)
and there will be a central ignition that results in a SN Ia.
We also repeated the calculation for $\chi=1$, namely with the disk wind having a zero terminal velocity.
In this case we find that higher circumbinary disk mass (larger $q_{d,0}$) is required for the companion WD to merge with the post-AGB core
while the later still hot.

This means that stars with massive envelopes ($M_{\rm{env}} \gtrsim 5~\rm{M_{\odot}})$ that engulf a WD companion,
might end the CE phase with a very close WD-WD binary.
This is in agreement with the results of \cite{LivioRiess2003}, who found that for a merger
to occur the AGB must be massive.
The WDs might merge before the hot WD (that was the post-AGB core) had time to cool much.
We note that as $\tau_{GW}$ is short, the probability of observing such systems is low.
But, observationally, there are indeed post-CE systems that have periods much shorter than simulations predict (\citealt{DeMarco2008}, 2009).
Our result that $\alpha_{\rm CE} < 1$ can account for this discrepancy.

It is in place to mention that recent studies also reach the conclusion that $\alpha_{\rm CE} < 1$.
\cite{DeMarco2011} perform simulations for many sets of binary parameters (see also \citealt{Passy2011}).
They find that there is a possible negative correlation between the mass ratio of the two stars and the value of $\alpha_{\rm CE}$.
As \cite{DeMarco2011} include the thermal energy (which is half of the binding energy according to the virial theorem), as we do here,
they also find that the values of $\alpha_{\rm CE}$ are below 1 (in cases where the thermal energy is not included, the
values of $\alpha_{\rm CE}$ they obtain are larger than 1).
\cite{Zorotovic2010} also come into the conclusion that $\alpha_{\rm CE} < 1$.
Their statistical analysis of observed post-CE binaries give $\alpha_{\rm CE} = 0.2$--$0.3$, but contrary to \cite{DeMarco2011} they do not
find any correlation between $\alpha_{\rm CE}$ and the mass of the companion.

\section{SUMMARY}
\label{sec:summary}

We conducted an analytical study of the ejection of a common envelope (CE) under some simplifying assumptions.
Despite the many 3D numerical hydrodynamical simulations of CE evolution, such an approach has its own merit,
as it points to the basic physics behind the finding that some of the envelope remains bound to the system.

Most of the binary gravitational energy in a CE is released in a short time during the final stages of the CE phase.
Here we took this time to be much shorter than the dynamical time in the outer regions of the envelope.
Therefore, the energy release process resembles a blast wave propagating from the center outwards.
We discuss two types of blast waves (section \ref{sec:blast}), which practically lead to the same result.
In the first type the energy release is instantaneous, i.e. an explosion, while in the second type there is a continuous energy deposition
over a short time.
We used a self similar solution to track the blast wave in each type as it propagates through the AGB envelope.
We found that part of the ejected envelope stays bound to the binary system, i.e., it does not reach the escape velocity.
This material is expected to fall back towards the center.
It is very likely that due to angular momentum conservation and further interaction of the fallback gas with the binary system,
a circumbinary disk will be formed.
The spherically symmetric geometry of the self similar solution does not allow the introduction of angular momentum to the solution of the blast wave.

As discussed in section \ref{sec:implications:disk}, the binary semi-major axis decreases
rapidly during the short circumbinary disk -- binary interaction period.
During that time the power of the system greatly exceeds the Eddington limit.
Most of the energy goes to blow the wind, but some goes to radiation.
It is therefore possible that such systems will be observed by the transient increase in their luminosity.

We showed that interaction of the binary system with the circumbinary disk will further reduce the orbital separation.
Consequently, in the alpha-prescription, we find that effectively $\alpha_{\rm CE} < 1$.
In many cases a merging will occur immediately after the dynamical phase of the CE.
In many cases where the companion is a WD, it will form with the remnant post-AGB core of the AGB (or RGB) core a WD-WD system
with very small separation.
In cases where the mass of the circumbinary disk is large enough, we suggest that
a core-degenerate (CD) is likely to occur --
an early merger of the WD and the post-AGB core will merge while the post-AGB core is still hot.
Such a merger is more likely to explode later as a SN Ia if the mass is above critical
and the AGB's core is more massive than the WD companion (see section \ref{sec:implications}).
The super-critical mass WD is stabilized by the rapid rotation.
Only after it slows down it will explode as a SN Ia.
We speculate, based on the results of \cite{YoonLanger2005}, that this time can be as long as
$\gtrsim {\rm few}\times 10^9 \yr$, and that the core-degenerate scenario is another channel leading to a SN Ia.

Our finding that effectively $\alpha_{\rm CE} < 1$ (see also \citealt{Ivanova2011}), can explain the recent findings of \cite{DeMarco2011}.
\cite{DeMarco2011} find that the value of $\alpha_{\rm CE}$ they deduce from observations is much smaller than
what their numerical simulations of the CE phase give (\citealt{DeMarco2008}, 2009, 2011).
We strongly encourage numerical study of the CE phase to follow the gas and its angular momentum after ejection
in order to find the amount of gas that falls back , and whether a circumbinary disk is formed.

\section*{ACKNOWLEDGEMENTS}
We thank Natalia Ivanova and Orsola De Marco for enlightening discussions,
and an anonymous referee for very helpful comments.
This research was supported by the Asher Fund for Space Research at the Technion and the Israel Science Foundation.
AK acknowledges a grant from the Irwin and Joan Jacobs Foundation.

\end{document}